\documentclass[aps,floatfix,lengthcheck,showpacs,amssymb,amsmath,amsfonts,
twocolumn ,nofootinbib,prd]{revtex4-1}
\usepackage[utf8]{inputenc}
\usepackage{color}
\usepackage{graphicx}
\usepackage{float}
\usepackage{subfigure}
\usepackage{acronym}
\usepackage{multirow}
\usepackage{tikz}
\usetikzlibrary{arrows,shapes,trees,decorations.pathreplacing}
\usepackage{hyperref}
\hypersetup{
    colorlinks=true,
    linkcolor=blue,
    filecolor=magenta,      
    urlcolor=cyan,
}
\usepackage[toc, nonumberlist]{glossaries}
\glsdisablehyper
\usepackage{verbatim}
\usepackage[binary-units=true, alsoload=astro]{siunitx}
\usepackage{bm}
\usepackage{mathtools}
\usepackage{rotating}
\usepackage[toc, nonumberlist]{glossaries}
\usepackage{multirow}

\graphicspath{images/}

\DeclarePairedDelimiterX\Basics[1](){ #1}

\begin{document}

\newglossaryentry{gw}{name=GW, description={Gravitational wave}}
\newglossaryentry{bbh}{name=BBH, description=Binary black hole}
\newglossaryentry{snr}{name=SNR, description=Signal to noise ratio}
\newglossaryentry{psd}{name=PSD, description=Power spectral density}
\newglossaryentry{elu}{name=ELU, description=Exponential linear unit}
\newglossaryentry{far}{name=FAR, description=false-alarm rate}
\newglossaryentry{fap}{name=FAP, description=false-alarm probability}
\newglossaryentry{usr}{name=USR, description=unbounded Softmax replacement}

\title[]{
Training Strategies for Deep Learning Gravitational-Wave Searches
}

\author{
    Marlin B. Sch{\"a}fer$^{1,2}$,
    Ond{\v{r}}ej Zelenka$^{3,4}$,
    Alexander H. Nitz$^{1,2}$,
    Frank Ohme$^{1,2}$,
    Bernd Br{\"u}gmann$^{3,4}$
}
\address{$^1$Max-Planck-Institut f{\"u}r Gravitationsphysik,
         Albert-Einstein-Institut, D-30167 Hannover, Germany}
\address{$^2$Leibniz Universit{\"a}t Hannover, D-30167 Hannover, Germany}
\address{$^3$Friedrich-Schiller-Universit{\"a}t Jena, D-07743 Jena, Germany}
\address{$^4$Michael Stifel Center Jena, D-07743 Jena, Germany}

\begin{abstract}
Compact binary systems emit gravitational radiation which is potentially detectable by current Earth bound detectors. Extracting these signals from the instruments' background noise is a complex problem and the computational cost of most current searches depends on the complexity of the source model. Deep learning may be capable of finding signals where current algorithms hit computational limits. Here we restrict our analysis to signals from non-spinning binary black holes and systematically test different strategies by which training data is presented to the networks. To assess the impact of the training strategies, we re-analyze the first published networks and directly compare them to an equivalent matched-filter search. We find that the deep learning algorithms can generalize low signal-to-noise ratio (\gls{snr}) signals to high \gls{snr} ones but not vice versa. As such, it is not beneficial to provide high \gls{snr} signals during training, and fastest convergence is achieved when low \gls{snr} samples are provided early on. During testing we found that the networks are sometimes unable to recover any signals when a false alarm probability $<10^{-3}$ is required. We resolve this restriction by applying a modification we call unbounded Softmax replacement (\gls{usr}) after training. With this alteration we find that the machine learning search retains $\geq 91.5\%$ of the sensitivity of the matched-filter search down to a false-alarm rate of $1$ per month.
\end{abstract}

\maketitle

\section{Introduction}
The direct detection of a gravitational wave (\gls{gw}) on September 14, 2015 \cite{gw150914} started the era of \gls{gw} astronomy. After the analysis of two and a half observing runs, tens of \gls{gw}s have been confirmed \cite{gwtc2, 3ogc}. GW170817 \cite{gw170817} was the first \gls{gw} event also seen in the electromagnetic spectrum \cite{Goldstein:2017, Savchenko:2017, Soares-Santos:2017, GBM:2017lvd}.

The latency between a \gls{gw} and its reported detection is a vital aspect of multi-messenger missions. Lowering the delay between data aggregation and signal detection allows to maximize electromagnetic observation time and reduces the risk that early emissions are being missed.

To extract \gls{gw} signals from the instrument data, a well-established technique known as \textit{matched filtering} is used in many search algorithms. It convolves \textit{templates}, i.e.\ pre-calculated models of the expected signals, with the measured data \cite{ligo_pipelines, gstlal_pipeline, mbta_pipeline, pycbc_live, spiir_pipeline}. When one of these templates matches the data to a given degree and the data quality is high enough, these searches report a candidate detection. 

Matched filtering is known to be optimal in stationary Gaussian noise when accurate models of the waveform exist~\cite{Allen:2005fk}. However, it can be computationally limiting when many templates are required. This is the case when effects such as higher-order modes \cite{Harry:2017:highermodes}, precession \cite{Harry:2016:precessing} or eccentricity \cite{Nitz:2019:eccentric} are considered. Furthermore, signals which are not covered by the filter bank may be missed entirely. While there are unmodeled searches that detect coincident excess power in different detectors \cite{cwb1, cwb2, olib}, they are less sensitive in regions where accurate models exist.

Recently, new deep learning based searches have started to be explored \cite{Huerta:2018a, Huerta:2018b, Gabbard:2018, Dreissigacker:2019, Krastev:2020, Schaefer:2020}. Summaries of the current state of the field are given in \cite{Cuoco:2020, huerta:2021}. The pioneering works by George et al.\ \cite{Huerta:2018a} and Gabbard et al.\ \cite{Gabbard:2018} demonstrated that deep neural networks are capable of detecting \gls{gw}s from two merging black holes (\gls{bbh}). The networks have also proven to generalize to signals with previously unseen parameters \cite{Huerta:2018a, Xia:2021}. It was shown that these algorithms can distinguish data containing a \gls{gw} from pure noise as well as matched filtering with a false-alarm probability (\gls{fap}) down to $10^{-3}$. That means, the networks were tested down to a level at which about 1 in 1000 pure noise samples was falsely classified as containing a signal.

The authors of \cite{Gebhardt:2019} find that the \gls{fap}s determined by the original studies do not directly translate to false-alarm rates (\gls{far}s) on continuous data streams. For \gls{far}s, the appropriate question to ask is how many false signals does the network identify per time interval of continuous data, as opposed to how many uncorrelated data chunks are falsely identified as containing a signal. The effects of clustering subsequent outputs when the network is applied via a sliding window have to be accounted for. Comparing deep learning searches to traditional matched-filter searches is, therefore, not trivial because matched-filter searches typically operate at \gls{far}s that are orders of magnitude smaller than what has been tested for early neural networks. In \cite{Schaefer:2020} we suggested a standardized testing procedure which produces statistics which are comparable to traditional search algorithms to resolve these issues.

In this paper we reanalyze and extend the results given in the initial papers \cite{Huerta:2018a, Gabbard:2018}. Our motivation is twofold. First, we want to verify and test the performance of the networks quoted in those papers. Specifically, we apply the testing procedure outlined in \cite{Schaefer:2020}. Second, we discuss how the \gls{gw} data is prepared for and presented to the network. The form of data preparation is often taken as a given, while comparatively more work is invested in finding a network structure that suits the problem. We carefully examine the influence of different choices of data presentation and training strategies on the ability to detect signals given a fixed network.

Here we focus on signal detection. The problem of deep learning parameter estimation is another vital and active area of research. Multiple groups have made advancements in this field \cite{Chua:2020, gabbard2020bayesian, Green:2020, Williams:2021, Schmidt:2020, Huerta:2018a}.

We use the network presented by Gabbard et al.\ \cite{Gabbard:2018} for most of our studies. It is trained on simulated data containing \gls{gw}s from \gls{bbh}s with individual black hole masses ranging from \SIrange{10}{50}{M_\odot}. The search is restricted to a single detector. This restriction reduces the parameter space to the two component masses, the orbital phase, the distance to the source and the time of coalescence.

The network classifies segments of \SI{1}{\second} duration sampled at \SI{2048}{\hertz} into the two categories ``$\text{noise}+\text{signal}$'' and ``noise'' by returning a value between 0 and 1 we call ``p-score''. A larger p-score corresponds to a higher confidence of the network that the input contains a signal.

To optimize the training strategy, we focus on the difference between \textit{curriculum learning} \cite{Bengio:2009} and \textit{fixed interval training}. Fixed interval training uses a single training set, i.e.\ a single, fixed range of signal-to-noise ratios (\gls{snr}s). Curriculum learning lowers the \gls{snr} of the training signals progressively, thus increasing the complexity with time. We evaluate different variants of both strategies. In total 15 different approaches are tested.

Each strategy is applied to 50 randomly initialized networks. We do this to guard against favorable initializations. All tests are done with two different implementations, to further increase robustness of our results. The different implementations use the two core libraries Tensorflow \cite{tensorflow_software} and PyTorch \cite{pytorch_software}, respectively.

We find that most training strategies are capable of closely reproducing the results given in \cite{Gabbard:2018}. We do not see a significant difference in performance between curriculum learning and fixed interval training strategies. However, networks that had access to lower \gls{snr} signals during training generally outperformed those that only saw high \gls{snr} signals. We find that networks trained on fainter signals can generalize to loud ones, while the opposite is not the case.

Further analysis of the networks showed that the efficiency, which is the fraction of correctly classified input samples containing a signal at a given \gls{fap}, drops to zero beyond a \gls{fap} of $10^{-3}$ when the training is carried out for long enough. This drop is caused by numerical instabilities in the final activation and the comparatively low penalty of false positives. We propose a simple modification that does not require retraining of the network to push this problem to significantly lower \gls{fap}s. We call this modification unbounded Softmax replacement (\gls{usr}).

We evaluate 3 different networks of each training strategy on a month of simulated data. The networks are applied using a sliding window with step size of \SI{0.1}{\second}. We follow the procedure outlined in \cite{Schaefer:2020} to analyze the results. Our evaluation of the base line network is limited by $\mathcal{O}\left(10^3\right)$ false alarms estimated with perfect confidence to contain a signal. By applying the \gls{usr} modification we are able to eliminate this restriction and can calculate sensitivities down to a \gls{far} of $1$ per month. For comparison, we construct a template bank and use it to do a matched-filter search on the same data used to evaluate the networks. We find that the machine learning search retains at least $91.5\%$ of the sensitivity of a matched-filter search for all tested \gls{far}s and most strategies.

All code required to reproduce our analysis is public and can be found at \cite{ml-training-strategies-github}.

The contents of this paper are structured as follows. In \autoref{sec:methods} we describe the architecture, data sets, training strategies, and evaluation methods. We apply these in \autoref{sec:results} and describe our findings. In particular we describe the \gls{usr} modification which allows the networks to be tested at low \gls{fap}s. We conclude in \autoref{sec:conclusion}.

\section{Methods}\label{sec:methods}

\subsection{General setup}\label{sec:methods:general}
We focus our studies on the network presented by Gabbard et al.\ in \cite{Gabbard:2018}. They used a convolutional neural network with 6 stacked convolutional layers followed by 3 fully connected layers. All but the last layer use an exponential linear unit (\gls{elu}) as activation function.

The architecture is altered in two details compared to the original version of \cite{Gabbard:2018}. We added a batch normalization layer before the first convolutional layer to take care of \textit{input normalization}. Input normalization scales all inputs to have a mean-value of 0 and a variance of 1. This is standard practice in contemporary deep learning and has been proven to help the network train efficiently \cite{Ioffe:2015}. The second modification is a reduction of the pool sizes. This change was required because we lowered the sample rate of the data from \SI{8192}{\hertz} to \SI{2048}{\hertz}. We decided to lower the sample rate for multiple reasons. First of all, the detector sensitivity drops sharply above \SI{1}{\kilo\hertz}. Thus little to no \gls{snr} is lost by disregarding higher frequencies. For this reason current searches are often limited to the same frequency band as well \cite{pycbc_live, Abbott:2020:grb, Abbott:2015:allsky}. Second, signals within our training set merge at much lower frequencies and do not exceed \SI{1}{\kilo\hertz}. Finally, as we will show in \autoref{sec:results}, our training converged to the same state as previous works. We are thus confident that this reduction in the sample rate has no negative impact on the network's ability to detect signals. A reduction of the size of the input to a neural network usually also helps with training. The resulting network setup is depicted in Table \ref{tab:network}.

\begin{table}[]
    \caption{\label{tab:network} The modified neural network from \cite{Gabbard:2018} as used in this study. The given shapes correspond to the tensor shapes in the TensorFlow version of the code, i.e.\ data length $\times$ number of channels. PyTorch swaps these dimensions. The order of the layers is given by reading the column ''layer type'' from top to bottom and left to right. Layers are grouped by their influence on the output shape and by trainable weights.
    }
    \begin{ruledtabular}

\iftrue
\begin{tabular}{lrr}
    layer type & kernel size & output shape \\
    \hline
    Input + BatchNorm1d & & $2048\times 1$ \\
    Conv1D + ELU & 64 & $1985\times 8$ \\
    Conv1D & 32 & $1954\times 8$ \\
    MaxPool1D + ELU & 4 & $488\times 8$ \\
    Conv1D + ELU & 32 & $457\times 16$ \\
    Conv1D & 16 & $442\times 16$ \\
    MaxPool1D + ELU & 3 & $147\times 16$ \\
    Conv1D + ELU & 16 & $132\times 32$ \\
    Conv1D & 16 & $117\times 32$ \\
    MaxPool1D + ELU & 2 & $58\times 32$ \\
    Flatten & & $1856$ \\
    Dense + Dropout + ELU & & $64$ \\
    Dense + Dropout + ELU & & $64$ \\
    Dense + Softmax & & $2$
\end{tabular}
\fi

    \end{ruledtabular}
\end{table}

All studies presented in \autoref{sec:methods:network_performance} and \autoref{sec:methods:training_strategies} were carried out using the network from George et al.\footnote{We adjusted the network from George et. al.\ too, by using batch normalization for input normalization and reducing the sample rate of the input to \SI{2048}{\hertz}.}\ \cite{Huerta:2018a} as well. However, with our particular training setup, every metric showed performance similar to the network from \cite{Gabbard:2018}. We present only results using the network based on the work of Gabbard et al.

Each \gls{gw} signal is defined by the component masses $m_1, m_2$ and a phase $\phi_0$. Two masses are drawn independently from a uniform distribution between \SI{10}{M_\odot} and \SI{50}{M_\odot} and the higher and lower values are assigned to $m_1$ and $m_2$, respectively, to enforce the condition $m_1\geq m_2$. Phases are uniformly drawn from the interval $\left[0,2\pi\right]$. We generate signals with 5 different phases for each pair of masses $\left(m_1,m_2\right)$. The amplitude, and therefore the distance of the source, is determined by the target \gls{snr} we have chosen. We fix the sky position to be overhead the LIGO Hanford detector \cite{aligo} and the inclination as well as the polarization to $0$, because in the case of nonprecessing signals and assuming a single detector, any variation in those parameters can be fully absorbed by modifications of the amplitude and phase of the signal.

The waveforms are generated with a sample-rate of \SI{2048}{\hertz}, a lower-frequency cutoff of \SI{20}{\hertz}, and using the model \verb|SEOBNRv4_opt| \cite{Devine:2016ovp} (optimized version of \verb|SEOBNRv4| \cite{Bohe:2016}). It is common practice to shift the location of the maximum amplitude by some small time within each training sample. This procedure allows the network to be less sensitive to the exact alignment of the waveform within its input. To achieve this behavior, we shift the position of the merger by a time uniformly drawn from \SIrange{-0.1}{0.1}{\second} before projecting onto the Hanford detector. After the projection the signals are \textit{whitened} using the analytic model of LIGO's design sensitivity at its zero detuned high power configuration \cite{lalsuite}, i.e., we divide the Fourier transformed signal by the square root of the power spectral density (\gls{psd}) associated with the power of the background noise at different frequencies, and transform back to the time domain. Whitening the data reduces the power at frequencies where the detector is known to be less sensitive. Next, the waveforms are scaled to an optimal \gls{snr} of 1. The optimal \gls{snr} $\rho_\text{opt}$ is defined by
\begin{equation}
    {\rho_\text{opt}}^2 = 4 \text{Re}\left[\int df \frac{\tilde{h}(f)\tilde{h}^\ast(f)}{S_n(f)}\right],
\end{equation}
where $\tilde{h}$ is the Fourier transform of the time domain signal, before it was whitened, $\tilde{h}^\ast$ is its complex conjugate, $S_n$ is the \gls{psd} and $\text{Re}$ extracts the real part of the complex number. Finally, we extract a time slice such that the original, not shifted merger time is located \SI{0.7}{\second} from the start of the window.

All noise is simulated from the same \gls{psd} used to whiten the signals. After generation, the noise is whitened by the \gls{psd} used to create it in the same way the signals are whitened. We choose to explicitly whiten the colored noise to take into account any artifacts the process may introduce. This also eliminates sources of errors and is in principle extendable to real noise.

The whitened signals and noise samples are combined during training. This allows us to rescale the signals at runtime to a desired strength. Since during generation all signals are scaled to \gls{snr} $1$, rescaling is achieved by a multiplication of the signal with the target \gls{snr}.

We have briefly tested training on frequency domain data. This was motivated by studies such as \cite{Dreissigacker:2019, Wei:2020}. While these studies analyze longer duration signals, there is no conceptual problem to using the frequency representation of short \gls{bbh} waveforms. To accommodate the complex valued frequency representation we changed the input layer in \autoref{tab:network} to a shape of $1025\times 2$ and inserted the real and imaginary parts as different channels. With this being our only modification to the architecture, the network was able to differentiate data containing a signal from pure noise but found up to $65\%$ fewer signals at low \gls{snr}. We suspect that with greater effort in finding an optimized architecture, one could regain the performance of the network on time domain data.

We have also explored training on raw data, i.e.\ data where the computationally expensive whitening is skipped. Even after several hundred epochs the network was not able to distinguish data containing signals from pure noise, irrespective of using the time or frequency domain representation of the data.

Our training set contains $20\,000$ unique combinations of component masses each of which is used to generate 5 waveforms with random coalescence phases. Therefore, it contains $100\,000$ individual signals. We generate $200\,000$ independent noise samples, $100\,000$ of which are used in combination with the signals. The remaining $100\,000$ noise samples are used as pure noise. Our training set, therefore, contains $200\,000$ independent samples.

The validation set is assembled in the same way as the training set. It too contains $100\,000$ samples of the ``signal''-class and $100\,000$ samples of the ``noise''-class for a total of $200\,000$ samples. The validation set was chosen to be of equal size to the training set due to its influence when curriculum training strategies are used. The conditions for when the complexity of the training set is increased are evaluated on this set.

We use a third data set to calculate relevant metrics of the network during training. This third set is required, as the validation set directly influences the training for curriculum strategies. Metrics determined on the validation set may, therefore, be biased. We call this third set the efficiency set and describe its usage in \autoref{sec:methods:network_performance}. It contains $10\,000$ unique signals and $400\,000$ independent noise samples.

Finally, we evaluate the performance of the network on a test set. This test set contains a month of simulated Gaussian noise with injections separated by a time uniformly distributed in the interval \SI[parse-numbers=false]{[16, 22]}{\second}. The injection parameters are drawn from the distributions shown in \autoref{tab:test_set}. Noise is generated using the same \gls{psd} used for the training set. A month of data corresponds to $\sim 26$ million correlated samples.

\begin{table}
    \caption{Injection parameters for the data set used to determine the \gls{far} and sensitive volume of the different networks.}
    \label{tab:test_set}
    \begin{ruledtabular}
    \begin{tabular}{lr}
        Parameter & Uniform distribution \\
        \hline
        Component masses & $m_1, m_2\in\ $\SI[parse-numbers=false]{\left(10, 50\right)}{M_\odot}\\
        Spins & 0\\
        Coalescence phase & $\Phi_0\in\left(0, 2\pi\right)$\\
        Polarization & $\Psi\in\left(0, 2\pi\right)$\\
        Inclination & $\cos{\iota}\in\left(-1, 1\right)$\\
        Declination & $\sin{\theta}\in\left(-1, 1\right)$\\
        Right ascension & $\varphi\in\left(-\pi, \pi\right)$\\
        Distance & \SI[parse-numbers=false]{d^2\in\left(500^2, 7000^2\right)}{{\mega\parsec}^2}
    \end{tabular}
    \end{ruledtabular}
\end{table}

Each network is trained for 200 epochs, i.e.\ 200 full passes on the training set. We found this to be a sufficient number of training cycles for most of the networks to converge to a stable performance on the validation set. We use the default implementations of the Adam optimizer with a learning rate of $10^{-5}$, $\beta_1 = 0.9$, $\beta_2 = 0.999$ and $\varepsilon = 10^{-8}$ \cite{adam}. As loss we use a variant of the binary crossentropy that is designed to stay finite,
\begin{equation}
    L(\bm{y}_\text{t}, \bm{y}_\text{p}) = -\frac{1}{N_\text{b}}\sum_{i=1}^{N_\text{b}} \bm{y}_{\text{t},i}\cdot\log\left(\epsilon + (1 - 2 \epsilon) \bm{y}_{\text{p},i}\right).
\end{equation}
Here $\bm{y}_\text{t}$ is either $(1, 0)^T$ for data containing a signal or $(0, 1)^T$ for pure noise, $\bm{y}_\text{p}$ is the prediction of the network, $N_\text{b}=32$ is the mini-batchsize, and $\epsilon = 10^{-6}$.

\subsection{Network performance}\label{sec:methods:network_performance}
A common metric when training neural networks is the \textit{accuracy}, which is the ratio of correctly classified samples over the total number of samples. This approach weighs false-negatives and false-positives equally.

\gls{gw} searches assign a statistical significance to each event. This is usually given as the \gls{far} of the search at the ranking statistic threshold associated with the candidate event. For the network we use the p-score as ranking statistic. The more false positives a search produces at a given ranking statistic, the less significant each event becomes. Therefore, false-positives severely limit the ability of the search to recover true events. Low latency searches do not distribute any event candidates publicly with a \gls{far} greater than $\sim1$ per month \cite{pycbc_live}. For searches which operate on archival data, low \gls{far}s are needed to assign a probability for the signal to be of astrophysical origin, based on the expected astrophysical rate of comparable events \cite{gwtc2,3ogc}.

For these reasons we monitor the \textit{efficiency} of the network rather than the accuracy. The efficiency is the true-positive probability at a fixed false-positive probability, i.e.,\ a fixed \gls{fap}. To do so, we sort the p-score outputs of the network on the noise from the efficiency set and use the $x$-th largest as a threshold, where we choose
\begin{equation}
    x = \lfloor N_n\cdot \text{\gls{fap}} \rfloor
\end{equation}
Here, $N_n$ is the total number of noise samples used and $\lfloor\cdot\rfloor$ denotes the flooring operation. We then evaluate the signals from the efficiency set scaled to \gls{snr}s $3, 6, 9, 12, 15, 18, 21, 24, 27$ and $30$ and count the samples that exceed the threshold. The efficiency is then given by
\begin{equation}
    \text{efficiency} = \frac{N_{s>t}}{N_s},
\end{equation}
with $N_{s>t}$ being the number of signals assigned a p-score larger than the threshold and $N_s$ the total number of signals. To get a better understanding of the efficiency as a function of the signal strength, we also calculate the efficiencies at each of the \gls{snr}s individually. In this work, a \gls{fap} of $10^{-4}$ is used for all efficiency calculations.

For each of the strategies we discuss in \autoref{sec:methods:training_strategies} the network is trained 50 times from scratch. The parameters of the networks are initially random for each run. The final performance of a single network may depend on these initial values. Training each network-strategy combination multiple times and averaging over their efficiencies reduces the influence of the network initialization, thus yielding greater insight into the impact of the training strategy.

After training has completed for all 50 networks, we choose 3 networks for which we calculate the sensitive volume and the false-alarm rate on a month of simulated data. The networks are chosen by the following scheme. We select the epoch of the maximum efficiency of all networks. At this epoch we pick the best and the worst performing networks, where ranking is based on the efficiency. The last network is chosen to be the one which has the efficiency closest to the average efficiency over all 50 runs at the chosen epoch.

The sensitivity and \gls{far} calculation follows the procedure outlined in \cite{Schaefer:2020}. As suggested in \cite{Huerta:2018a, Gabbard:2018}, the network is applied to time series data of duration longer than the input window via a sliding window. We choose a step size of \SI{0.1}{\s} to ensure the correct alignment of the merger time within the input window for at least one step. Each window is whitened individually using the same method and noise model applied to the training set.

To reduce the computational cost, the data are sliced into the input windows and preprocessed only once. We store this sliced data and apply the different networks to it. This allows us to evaluate the entire month of data in about \SI{1}{\hour} on a single NVIDIA RTX 2070 SUPER.

The network outputs a value between $0$ and $1$ for every slice. A value of $1$ corresponds to the network being confident that it has seen a signal. We use this output as ranking statistic. Outputs that exceed a threshold, which we call trigger-threshold, are clustered by their time. Within each cluster the first time where the output becomes maximal is picked. The combination of this time and the corresponding network output is called an event.

The list of events is compared to the known injection times. If the event is separated from the closest injection by more than some maximum time it is called a false positive. Otherwise we consider it a true positive. From these we can calculate the \gls{far} as well as the sensitive volume as detailed in \cite{Schaefer:2020}. The \gls{far} is given by
\begin{equation}
    \text{\gls{far}} = \frac{N_f}{T_o} ~,
\end{equation}
where $N_f$ is the number of false positives and $T_o$ is the duration of the analyzed data. When the injections are distributed uniformly in volume the sensitive volume of the search is given by
\begin{equation}
    V\left(\text{\gls{far}}\right) = V\left(d_\text{max}\right)\frac{N_t\left(\text{\gls{far}}\right)}{N_i},
\end{equation}
where $d_\text{max}$ is the maximum distance at which sources are injected, $V\left(d_\text{max}\right)$ is the volume of a sphere with radius $d_\text{max}$, $N_i$ is the total number of injections and $N_t\left(\text{\gls{far}}\right)$ is the number of true positives at a given \gls{far}. The \gls{far} can be adjusted by considering only events above a given threshold. To convert the sensitive volume to a distance we calculate the radius of a sphere of the given volume.

We use a p-score of $0.1$ as our trigger-threshold. Triggers are said to belong to a cluster if they are within \SI{0.2}{\second} of the cluster bounds. An event is called a true-positive if there was an injection within \SI{0.3}{\second} of the reported event time. Otherwise it is a false-positive. We chose the cluster boundary time as twice the step size to allow for modest smoothing of the network output, while keeping it short compared to the average duration of a signal ($\mathcal{O}\left(\text{\SI{1}{\second}}\right)$). The maximum separation between an event and the corresponding injection was chosen to be larger than the cluster boundaries but still small compared to the average signal duration. None of these parameters were optimized.

\begin{figure}
    \centering
    \includegraphics[width=0.45\textwidth]{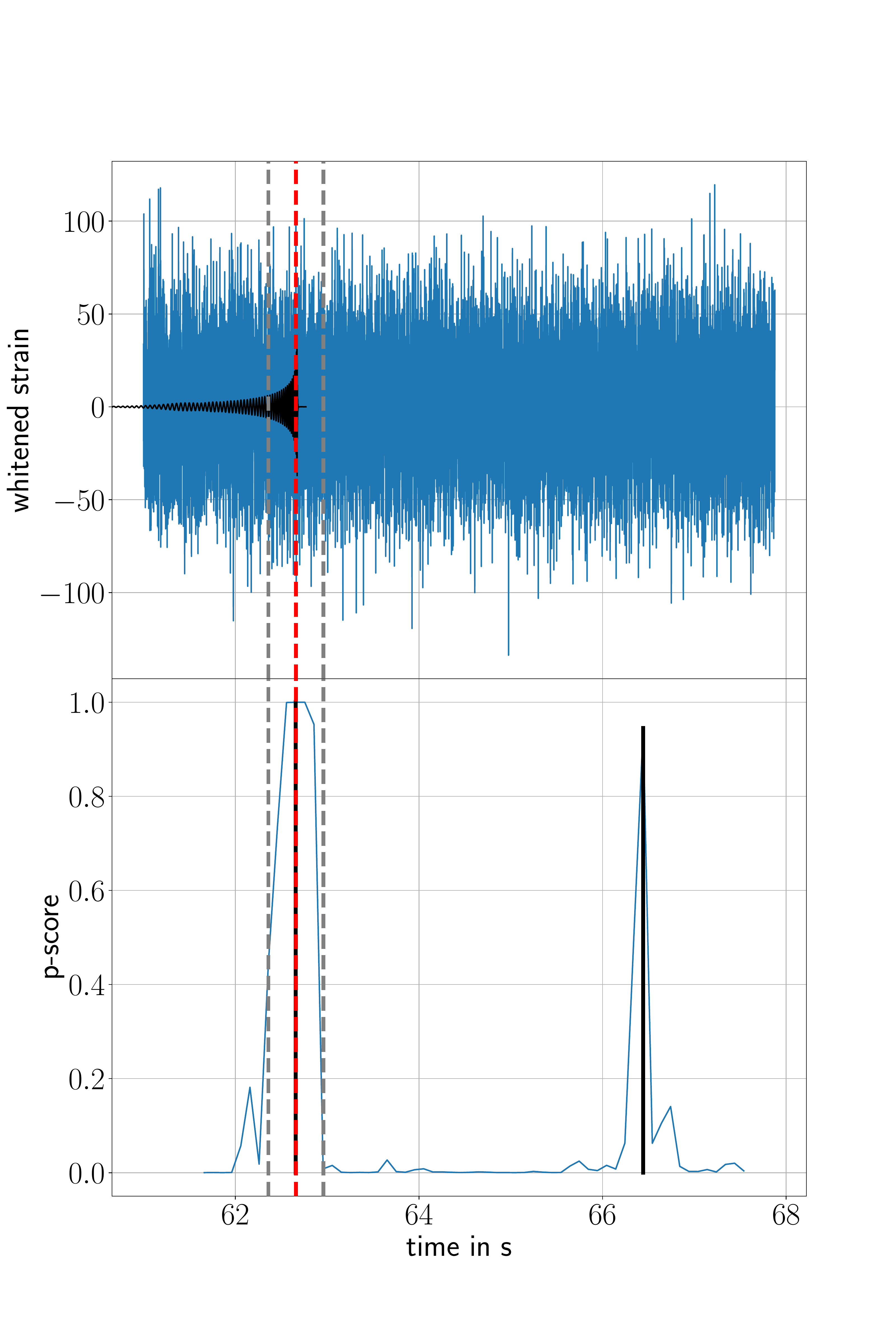}
    \caption{A sample output from the network on long duration data. The top panel shows the whitened input data. The injected signal is overlayed in black. The red vertical line signifies the time of the injection, i.e.\ the time that would ideally be returned by the search algorithm. The vertical grey lines mark the interval within which a returned event is classified as a true positive. The bottom panel shows the output of the network corresponding to the input. The vertical red and grey lines, again, show the true injection time and the allowed interval for true positives respectively. The black vertical lines mark the events returned by the search. Their height is the p-score attributed to the event. While the first event is a true positive, the second event is a false-positive originating from noise.}
    \label{fig:input_output}
\end{figure}

\autoref{fig:input_output} shows example output from one of the networks. The top panel shows the raw input with the injected waveform overlayed in black. The injection time is marked with a red vertical line and the grey lines highlight \SI[parse-numbers=false]{\pm 0.3}{\second} where events are true positives. The bottom panel shows the network output for the corresponding time. The black vertical lines show the events returned by the search.

\subsection{Training strategies}\label{sec:methods:training_strategies}
The two initial publications by George et al.\ \cite{Huerta:2018a} and Gabbard et al.\ \cite{Gabbard:2018} disagree on the usefulness of curriculum learning. Whereas George et al.\ find a noticeable improvement by using curriculum learning, Gabbard et al.\ find no difference in the final performance of the network.

We aim to determine the impact curriculum learning has on the final performance and speed of convergence of these networks. By doing so we optimize the sensitivity of the networks tested here and hope that our findings generalize also to state-of-the-art machine learning search algorithms \cite{Krastev:2020, Schaefer:2020, gabbard2020bayesian, Wei2021}.

Our study contains 10 curriculum learning and 5 fixed interval training strategies. An overview can be found in \autoref{tab:studies}. The minimum \gls{snr} allowed in any of these strategies is $\geq 5$. We choose \gls{snr} $5$ as a lower bound as this is roughly the lowest single detector \gls{snr} at which signals seen in multiple detectors can be confidently distinguished from noise \cite{gwtc2,3ogc}.

We test 5 different conditions for the optimal \gls{snr} contained in the training data for both types of strategies. For curriculum strategies, these conditions prescribe when the \gls{snr} of the training data is lowered. For fixed interval strategies the conditions are the interval from which the \gls{snr} for each sample is drawn.

Curriculum strategies use either the validation loss, the validation accuracy, or the number of epochs since the last step as conditions. For validation loss and validation accuracy we choose either a threshold or wait until the values stabilize and do not improve anymore. The latter are labeled by a prefix ''plateau'' throughout this paper. We choose a threshold of $0.95$ for the validation accuracy and $0.2$ for the validation loss. These values are arbitrary but proved to work well. When using the plateau conditions we lower the training range when the validation loss or validation accuracy, respectively, do not improve by more than $0.01\%$ for $6$ consecutive epochs. Finally, we also test lowering the training \gls{snr} irrespective of any of the metrics, by waiting 5 epochs between steps. We choose to wait $5$ epochs to allow the network enough time at each signal strength while ensuring we reach the minimum \gls{snr}. No extensive studies testing different values were made.

We test two different approaches to lowering the training \gls{snr}. All curriculum strategies start with \gls{snr}s which are uniformly drawn from the interval $\left[90, 100\right]$. Strategies that are given the postfix ''relative'' lower the bounds of this interval by $10\%$ at each step. The ranges are not lowered further when the lower bound of the interval reaches \gls{snr} $5$. Strategies without the postfix ''relative'' lower the bounds of the interval by a fixed value of $5$ at each step. This procedure is also continued down to a minimum bound of \gls{snr} $5$.

For fixed interval training strategies we test training on a single \gls{snr} as well as a fixed size interval of \gls{snr}s. We choose to train on fixed single \gls{snr}s $8$, $15$ and $30$ to cover the low, mid and high \gls{snr}s respectively. Training on a single \gls{snr} allows us to test how well the network generalizes to lower and higher \gls{snr}s than it has seen during training. By drawing the \gls{snr} from an interval we aim to reduce the dependence on a specific signal strength. We choose two strategies that draw \gls{snr}s from a fixed range. One covers only the lowest range used by any of the non-relative curriculum strategies, i.e.\ it draws the signal \gls{snr}s from the interval $\text{\gls{snr}}\in\left[5,15\right]$. The other draws the \gls{snr}s from the entire range of \gls{snr}s seen by the curriculum strategies, $\text{\gls{snr}}\in\left[5,100\right]$.

\begin{table}
\caption{An overview of the different training strategies tested in this work. The ''Curriculum'' type strategies lower the \gls{snr} of the training samples whenever the condition in the last column is fulfilled. All of them start with  $\mathrm{SNR}\in\left[90, 100\right]$. Curriculum strategies with the postfix ''relative'' in their name lower the boundaries of the interval by $10\%$ at each step, until the lower limit falls below \gls{snr} 5. The other curriculum strategies lower the bounds by a fixed value of $5$, until the lower limit reaches \gls{snr} 5. A metric fulfills the plateau condition when it has not improved by more than $0.01\%$ for $6$ consecutive epochs. The ''Fixed interval'' type strategies use a single \gls{snr} range for the entire training. Their interval is given in the last column. }
\label{tab:studies}
\begin{tabular}{p{0.25\columnwidth}|p{0.35\columnwidth}|p{0.35\columnwidth}}
    \hline\hline
    \multicolumn{1}{l}{Type} & \multicolumn{1}{l}{Name} & \multicolumn{1}{l}{Condition} \\
    \hline
    \multirow{12}{*}{Curriculum} & accuracy & \multirow{2}{*}{\parbox[t]{2.4cm}{when validation\\ accuracy $\geq 0.95$}}\\
    \cline{2-2}
    & accuracy relative & \\
    \cline{2-3}
    & epochs & \multirow{2}{*}{every 5 epochs} \\
    \cline{2-2}
    & epochs relative & \\
    \cline{2-3}
    & loss & \multirow{2}{*}{\parbox[t]{2.4cm}{when validation\\ loss $\leq 0.2$}} \\
    \cline{2-2}
    & loss relative & \\
    \cline{2-3}
    & plateau accuracy & \multirow{3}{*}{\parbox[t]{3cm}{6 epochs validation\\ accuracy plateau}} \\
    \cline{2-2}
    & plateau accuracy & \\
    & relative & \\
    \cline{2-3}
    & plateau loss & \multirow{3}{*}{\parbox[t]{3cm}{6 epochs validation\\ loss plateau}} \\
    \cline{2-2}
    & plateau loss & \\
    & relative & \\
    \hline
    \multirow{5}{*}{Fixed interval} & \gls{snr} 30 & $\mathrm{SNR}=30$ \\
 \cline{2-3}
    & \gls{snr} 15 & $\mathrm{SNR}=15$ \\
    \cline{2-3}
    & \gls{snr} 8 & $\mathrm{SNR}=8$ \\
     \cline{2-3}
    & low & $\mathrm{SNR}\in \left[5, 15\right]$ \\
     \cline{2-3}
    & full & $\mathrm{SNR}\in \left[5, 100\right]$ \\
    \hline\hline
\end{tabular}
\end{table}

\subsection{Matched-filter baseline}\label{sec:methods:mf}
In order to assess how sensitive the trained networks are in relation to conventional searches, we perform a matched-filter analysis of the test set described in \autoref{sec:methods:general}. To do so, we utilize the PyCBC analysis toolkit \cite{pycbc-github}.

The template bank covers component masses from \SIrange{10}{50}{M_\odot} and is constructed to lose no more than $3\%$ of the \gls{snr} of any signal due to its discreteness. The templates are placed stochastically. In total, the bank contains $598$ templates.

The search is implemented by \verb|pycbc_inspiral|. We configured it to output a set of times where any template of the bank convolved with the data exceeds a matched-filter \gls{snr} of $5$. Unlike the optimal \gls{snr}, the matched-filter \gls{snr} is the match of a detector data segment with a template, and so it varies based on the noise realization, while the optimal \gls{snr} assumes a noise realization that is constant zero. Combining the times where the threshold is exceeded with the corresponding matched-filter \gls{snr} and by using this \gls{snr} as ranking statistic, we obtain a set of triggers. We then process these triggers as described in \autoref{sec:methods:network_performance} to find events and calculate \gls{far}s and sensitive distances.

The configuration files are included in the data release \cite{ml-training-strategies-github}.

\section{Results}\label{sec:results}
\subsection{Sensitivities}\label{sec:results:sensitivities}
\begin{figure}
    \centering
    \includegraphics[width=0.5\textwidth]{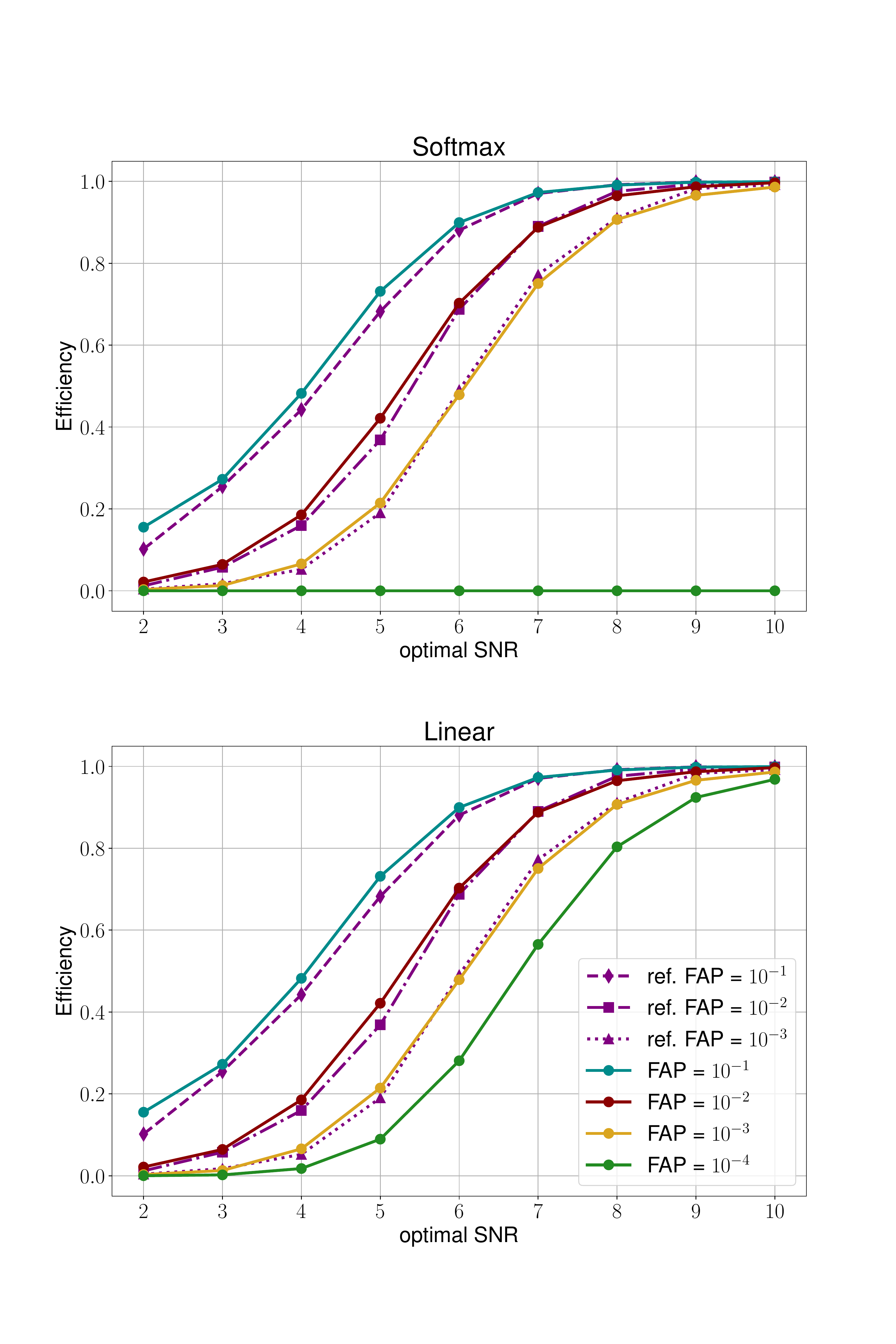}
    \caption{The efficiency as a function of optimal \gls{snr} at different \gls{fap}s. The network was trained on \gls{snr}s drawn from the fixed interval $\left[5,15\right]$. We used epoch $186$ of the network with the lowest efficiency at that epoch to produce this figure. The top panel shows the efficiency when the last layer uses a Softmax activation, the bottom panel shows the same network with the \gls{usr} modification. We determine the threshold on the network output using a set of $400\,000$ pure noise samples. Any of the $10\,000$ signals at each \gls{snr} exceeding this threshold are counted as detected. We compare our findings to Figure 3 of the reference \cite{Gabbard:2018} which closely reproduces efficiencies of matched filtering.}
    \label{fig:efficiency_example}
\end{figure}

We are able to reproduce or in some cases even improve on the results given in \cite{Gabbard:2018}. The top panel of \autoref{fig:efficiency_example} shows the efficiency of one network as a function of the \gls{snr} at fixed \gls{fap}s calculated on the efficiency set. We compare our findings to theirs and find excellent agreement with the results shown in Figure 3 of \cite{Gabbard:2018}, which closely reproduced efficiencies of matched filtering. The efficiencies at \gls{fap}s down to $10^{-3}$ for most other training strategies also closely follow the findings of Gabbard et al. We are, therefore, able to robustly reproduce the findings of \cite{Gabbard:2018}.

\begin{figure*}
    \centering
    \includegraphics[width=\textwidth]{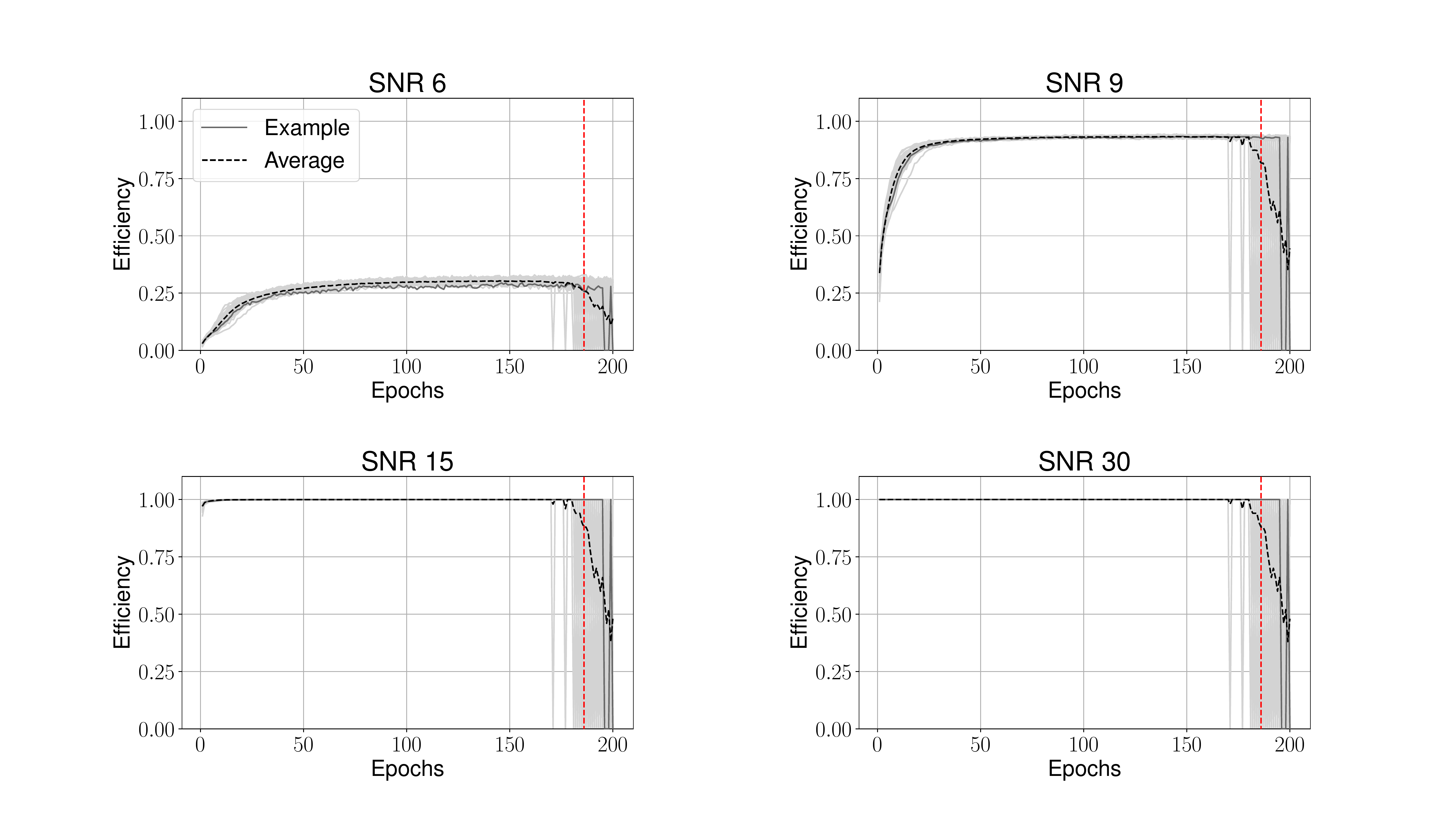}
    \caption{The evolution of the efficiency as a function of the epochs at different optimal \gls{snr}s. Training used the fixed \gls{snr} interval $\left[5,15\right]$. The individual evolutions of all $50$ runs are included as grey curves that form overlapping grey bands when plotted together. The dashed black line is the average of those. In dark grey we highlight the evolution of the efficiency for a single network. At the epoch marked by the red, dashed, vertical line we select the network with the highest, lowest and closest to average efficiency for further testing. The curves are computed at a \gls{fap} of $10^{-4}$.}
    \label{fig:efficiency_evolution_fixed_low}
\end{figure*}

In \autoref{fig:efficiency_evolution_fixed_low} we show the evolution of the efficiency of the $50$ networks trained on the fixed interval $\text{\gls{snr}}\in\left[5,15\right]$ as the number of training epochs is increased at a \gls{fap} of $10^{-4}$. Each panel of the plot shows the efficiency for a chosen \gls{snr} which allows us to observe how well the networks perform during different stages of the training at different signal strengths. This is especially interesting for curriculum strategies, where the \gls{snr} in the training set is adjusted as the network trains. The grey lines show the evolution of the efficiency for the different network initializations. The black, dashed line is the average of the grey lines. We highlight the evolution of a single network in dark grey. The red, dashed, vertical line signifies the epoch of maximum efficiency over all $50$ networks and $200$ epochs. 

All networks in \autoref{fig:efficiency_evolution_fixed_low} converge to similar efficiencies during the first $\sim 100$ epochs. However, as training continues sudden drops to zero efficiency occur which become more frequent at later epochs. As a result the average efficiency drops continuously after some time. All networks show this behavior and thus the influence of an unlucky initialization can be ruled out. Furthermore, the drops are observed at all \gls{snr}s simultaneously and, therefore, do not depend on the signal strength.

The same effect can be seen in the top panel of \autoref{fig:efficiency_example}. For $\text{\gls{fap}s}\geq 10^{-3}$ the curves behave as expected. As one lowers the \gls{fap} the efficiency at any given \gls{snr} is expected to drop. Visually this manifests in a shift of the efficiency curves toward higher \gls{snr}s. Ideally, this behavior would be true for any \gls{fap}. However, at a \gls{fap} of $10^{-4}$ the efficiency collapses and becomes a constant $0$.

The drops to zero efficiency are caused by noise samples which are attributed a p-score of 1. Since the Softmax activation on the last layer restricts outputs to the interval $\left[0, 1\right]$, no signal samples can achieve a p-score larger than the threshold and thus they cannot be distinguished from noise.

Many of the noise samples attributed a p-score of 1 are caused by numerical rounding errors in the Softmax activation 
\begin{equation}\label{eq:softmax}
    {\text{Softmax}\left(\bm{x}\right)}_i=\frac{\exp\left(x_i\right)}{\sum_{j=0}^N \exp\left(x_j\right)} ~,
\end{equation}
where $\bm{x} = \left(x_0, x_1, \hdots, x_N\right)$ is the vector of outputs of the previous layer in the network, and $N+1$ is the number of neurons in the layer.

The networks operate with single precision (32-bit) floating point numbers. Therefore, small changes in the values of $\bm{x}$ may cause a roundoff error due to the rapid change in scale of the exponential functions. When this occurs, the fraction may evaluate to $1$ even when mathematically \eqref{eq:softmax} may never be $1$.

We removed the final activation of the pre-trained network in an attempt to avoid the rounding errors. To do so, we recast \eqref{eq:softmax} for $N=1$ into
\begin{equation}\label{eq:recast_softmax}
    \frac{\exp\left(x_0\right)}{\exp\left(x_0\right)+\exp\left(x_1\right)}=\frac{1}{1+\exp\left(x_1-x_0\right)},
\end{equation}
and impose thresholds for the efficiency calculation on the difference $x_0-x_1$ directly rather than ${\text{Softmax}\left(\bm{x}\right)}_0$. Since \eqref{eq:recast_softmax} is bijective, there exists a direct relation between thresholds in $x_0-x_1$ and the thresholds on $\text{Softmax}\left(\bm{x}\right)_0$. We use $x_0-x_1$ rather than $x_1-x_0$ as our ranking statistic since $x_0-x_1>\hat{x}_0-\hat{x}_1\Leftrightarrow {\text{Softmax}\left(\bm{x}\right)}_0>{\text{Softmax}\left(\bm{\hat{x}}\right)}_0$. We call this modification unbounded Softmax replacement.

\begin{figure*}
    \centering
    \includegraphics[width=\textwidth]{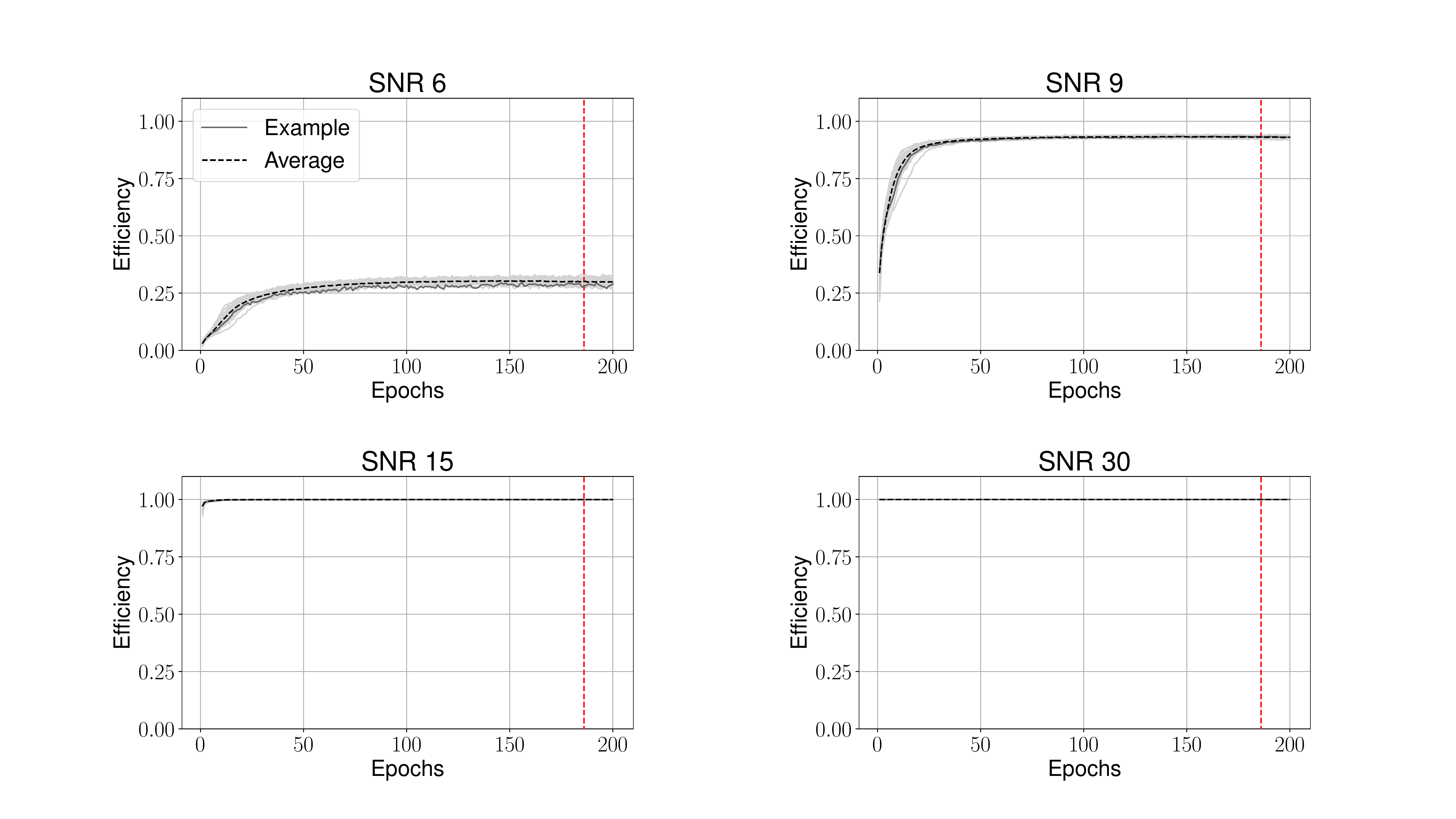}
    \caption{The evolution of the efficiency as a function of the epochs at different optimal \gls{snr}s. Training used the fixed \gls{snr} interval $\left[5,15\right]$. The individual evolutions of all $50$ runs are included as grey curves that form an overlapping grey band when plotted together. The dashed black line is the average of those. In dark grey we highlight the evolution of the efficiency for a single network. At the epoch marked by the red, dashed, vertical line we select the network with the highest, lowest and closest to average efficiency for further testing. The curves are computed at a \gls{fap} of $10^{-4}$. This figure shows the same networks as \autoref{fig:efficiency_evolution_fixed_low} after applying \gls{usr}. This prevents the efficiency to drop to 0.}
    \label{fig:efficiency_evolution_fixed_low_lin}
\end{figure*}

The resulting efficiency is depicted in the bottom panel of \autoref{fig:efficiency_example}. \autoref{fig:efficiency_evolution_fixed_low_lin} shows the efficiency evolution at different optimal \gls{snr}s. We find that the drops to zero efficiency vanish when we apply \gls{usr}. This is the case for all training strategies we explored and more examples are shown in the appendix (see \autoref{fig:efficiency_evolution_acc_rel_soft} to \autoref{fig:efficiency_evolution_fixed_30_lin}).

One could also try to resolve the rounding issue by using double precision (64-bit) floating point numbers instead of single precision when applying the Softmax layer. We have tested a numerically safe implementation of the Softmax and found that its first output is rounded up to one even for quadruple (128-bit) precision when the difference $x_0-x_1>45$. This is a relatively low value that indeed occurs for some noise realizations in our experiments. Although using higher precision for the Softmax layer increases the range of values it can operate on, the \gls{usr} still solves roundoff issues more robustly.

The efficiency is a metric that is easy to calculate and physically more relevant than the accuracy of the network. However, it does not deal with samples where waveforms are misaligned in the data or take into account longer stretches of time. It is, therefore, only an approximation to the true statistic we want to calculate: the sensitive volume.

To assess if the efficiency is a good approximation to this statistic, we calculate the sensitive volume of three chosen networks for every training strategy as described in \autoref{sec:methods:network_performance}. The networks are chosen from the $50$ different initializations based on their efficiency at a selected epoch. We pick the networks with the highest, lowest and closest to average efficiency and denote them with ''High'', ''Low'' and ''Mean'', respectively, from here on out. If the efficiency at a fixed \gls{fap} is a good indicator of the networks sensitivity we expect the sensitive volume to scale with the efficiency.

\begin{figure}
    \centering
    \includegraphics[width=0.48\textwidth]{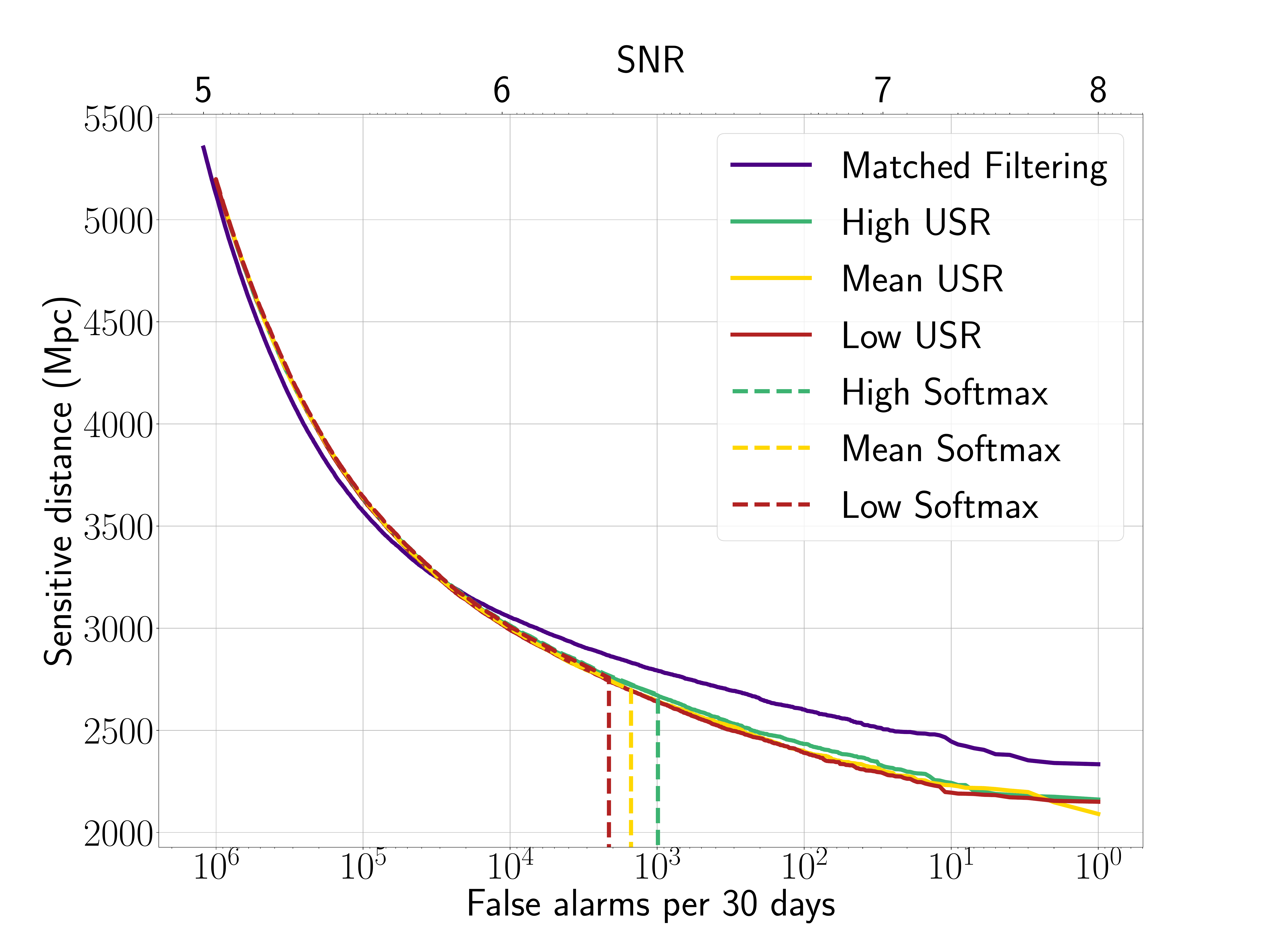}
    \caption{The sensitive distance as a function of the \gls{far} (bottom horizontal axis) for different search algorithms. We compare differently initialized networks trained on data containing signals with \gls{snr} $\in\left[5,15\right]$ to an equivalent matched-filter search. The dashed lines show the original networks, the filled lines show the corresponding network when \gls{usr} is applied. The labels ''High'' (green), ''Mean'' (yellow) and ''Low'' (red) correspond to the networks with the highest, closest to average and lowest efficiency at epoch $186$, respectively. In purple we show the equivalent matched-filter search that operates with a template bank containing $598$ templates. The top horizontal axis shows the \gls{snr} threshold for the matched-filter search corresponding to the \gls{far} on the bottom axis.}
    \label{fig:sensitivity_fixed_low}
\end{figure}

\autoref{fig:sensitivity_fixed_low} shows the sensitive distance as a function of the \gls{far} computed for the three networks trained on the fixed, low \gls{snr} interval. It compares the networks with (dashed) and without (continuous) the final Softmax activation and shows an equivalent matched-filter search in purple as reference. We find that the network is sensitive to sources up to a distance of \SI{2150}{\mega\parsec} with $1$ false alarm per month.

The sensitive radii of all converged deep learning searches lie within $3.4\%$ of each other for \gls{far}s where all of them are non-zero. However, the sensitivity of the networks with the final Softmax activation drops to zero for \gls{far}s $\leq\mathcal{O}(10^3)$ per month. This drop is caused by $\mathcal{O}(10^3)$ false alarms with a p-score of 1. This saturation of the final activation can be alleviated by applying the \gls{usr} modification and using the new output as a ranking statistic.

\begin{figure}
    \centering
    \includegraphics[width=0.48\textwidth]{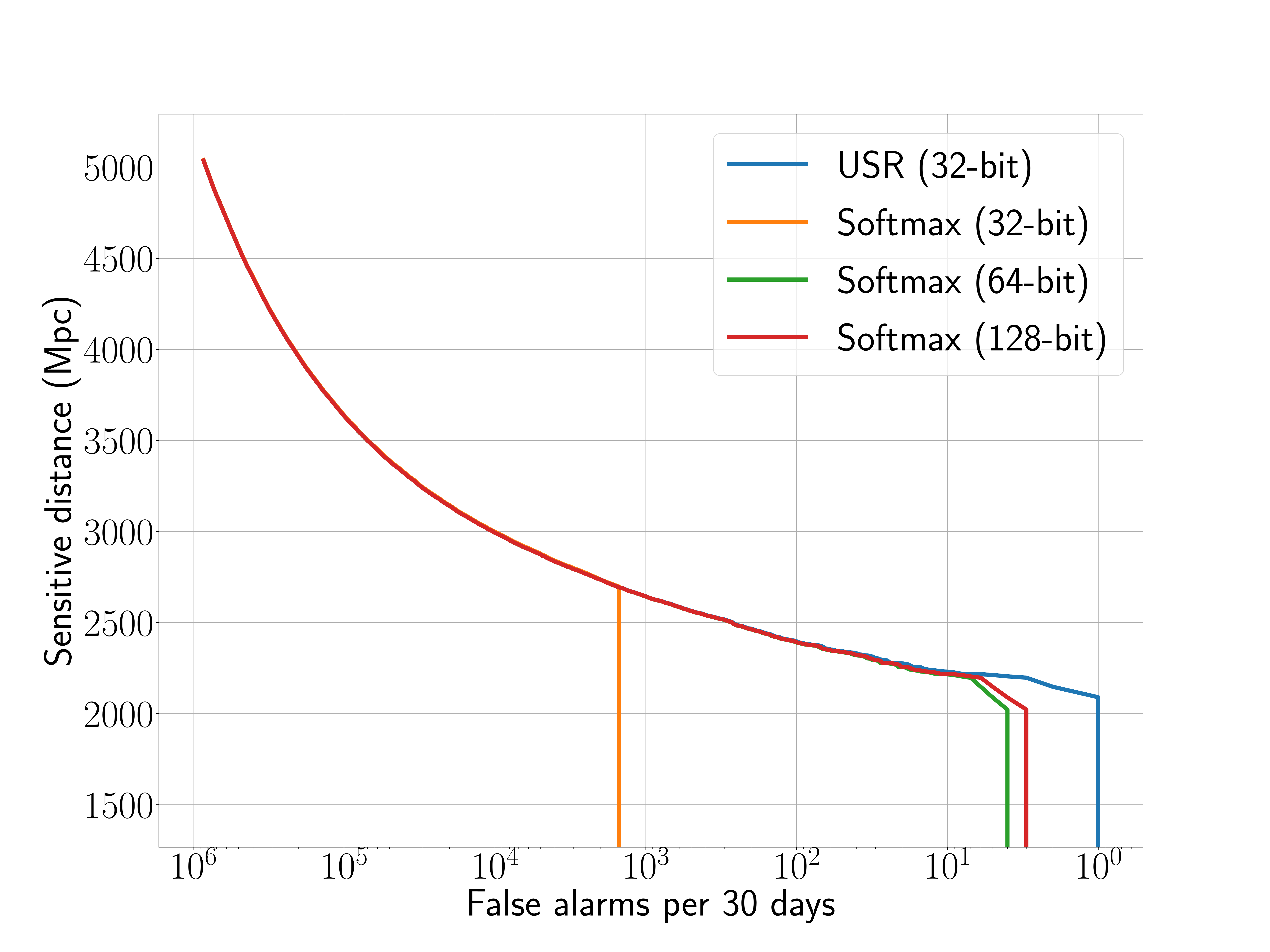}
    \caption{Sensitivity of the "mean" run of the "fixed low" strategy using the Softmax layer of various floating point precisions as well as the USR. A similar behavior of USR performing at least as well as the Softmax with all precisions was observed in all 45 evaluated runs.}
    \label{fig:precision_sensitivities}
\end{figure}

All tested networks have also been re-evaluated using higher precision floating point data types for the final activation function evaluation (example shown in Fig. \ref{fig:precision_sensitivities}). This resulted in the networks remaining sensitive at \gls{far}s down to $3$ per month. However, applying the \gls{usr} modification allowed us to test the network down to a \gls{far} of $1$ per month. Additionally, casting to a higher precision considerably increases computation time in the network due to hardware optimizations of GPUs for single precision floating point operations. In our view, the effectiveness of the \gls{usr} outweights the benefits of using higher precision, hence we only report results obtained with the \gls{usr} modification.

We had expected to find networks with higher efficiencies to be more sensitive, even within different initializations of the same training strategy. As such in \autoref{fig:sensitivity_fixed_low} we expected to find the sensitivity curve labeled ''High'' to be above the one labeled ''Mean'' above the one labeled ''Low''. While this is true in the example shown in \autoref{fig:sensitivity_fixed_low} in some regions, for other training strategies the order is arbitrary. All initializations converge to basically the same sensitivity. Sensitivity plots for all training strategies are provided in the data release \cite{ml-training-strategies-github}.

The machine learning algorithms are compared to an equivalent matched-filter search, shown in purple in \autoref{fig:sensitivity_fixed_low}. All searches perform equally well for \gls{far}s $\geq 10^5$ per month. For smaller \gls{far}s the matched-filter search is sensitive to sources which are up to \SI{200}{\mega\parsec} farther away. The deep learning search retains at least $91.5\%$ of the sensitivity compared to the matched-filter search at all \gls{far}s.

This result shows that with minimal modification to the architecture the original network from \cite{Gabbard:2018} achieves a sensitivity comparable to matched filtering for short \gls{bbh} signals in simulated Gaussian noise even at \gls{far}s previously untested for this particular network architecture.

All results above were obtained on data generated and whitened by the exact same \gls{psd} used during training. For realistic searches, this assumption does not hold as the \gls{psd} in the detectors drifts over time \cite{LIGOScientific:2021djp}. To assert that the network does not depend strongly on the exact \gls{psd} used during training, we also evaluated the sensitivity using a version of the training-\gls{psd} scaled by a constant factor of $1.05$ in all frequency bins. This reduced the sensitive distance at all \gls{far}s by roughly $1/\sqrt{1.05}$, in agreement with the theoretical expectation.

We also tested the effects of using a realistic variation of the \gls{psd}. To determine the variation, we used $20$ \gls{psd}s derived on real data from the O3a observing run \cite{LIGOScientific:2019lzm}, chose one as reference, and divided it by all the others. We then determined the PSD ratio that had the largest mean deviation from unity and multiplied it with the training \gls{psd} to obtain a realistically varied \gls{psd}. Generating and whitening the data by this varied \gls{psd} reduces the sensitivity at \gls{far}s below $100$ per month to around the same level as is observed for the scaled \gls{psd}.

Finally, we tested whitening by a different \gls{psd} than the one used for generating the data. For this purpose we used a second \gls{psd} variation to whiten the data generated by the first \gls{psd} variation described above. To obtain the second \gls{psd}, we used the PSD ratio that had the smallest, instead of the largest, mean deviation from unity. This simulates a worst-case scenario for realistic \gls{psd} variations. We find that the sensitivity drops by as much as $20\%$ compared to using the correct \gls{psd} for whitening. This analysis shows that the network is robust against differences between the training \gls{psd} and the \gls{psd} of the analyzed data, as long as the correct \gls{psd} is used for whitening.

\subsection{Training strategies}
\begin{figure*}
    \centering
    \includegraphics[width=\textwidth, trim=2.5cm 5cm 2.5cm 5cm]{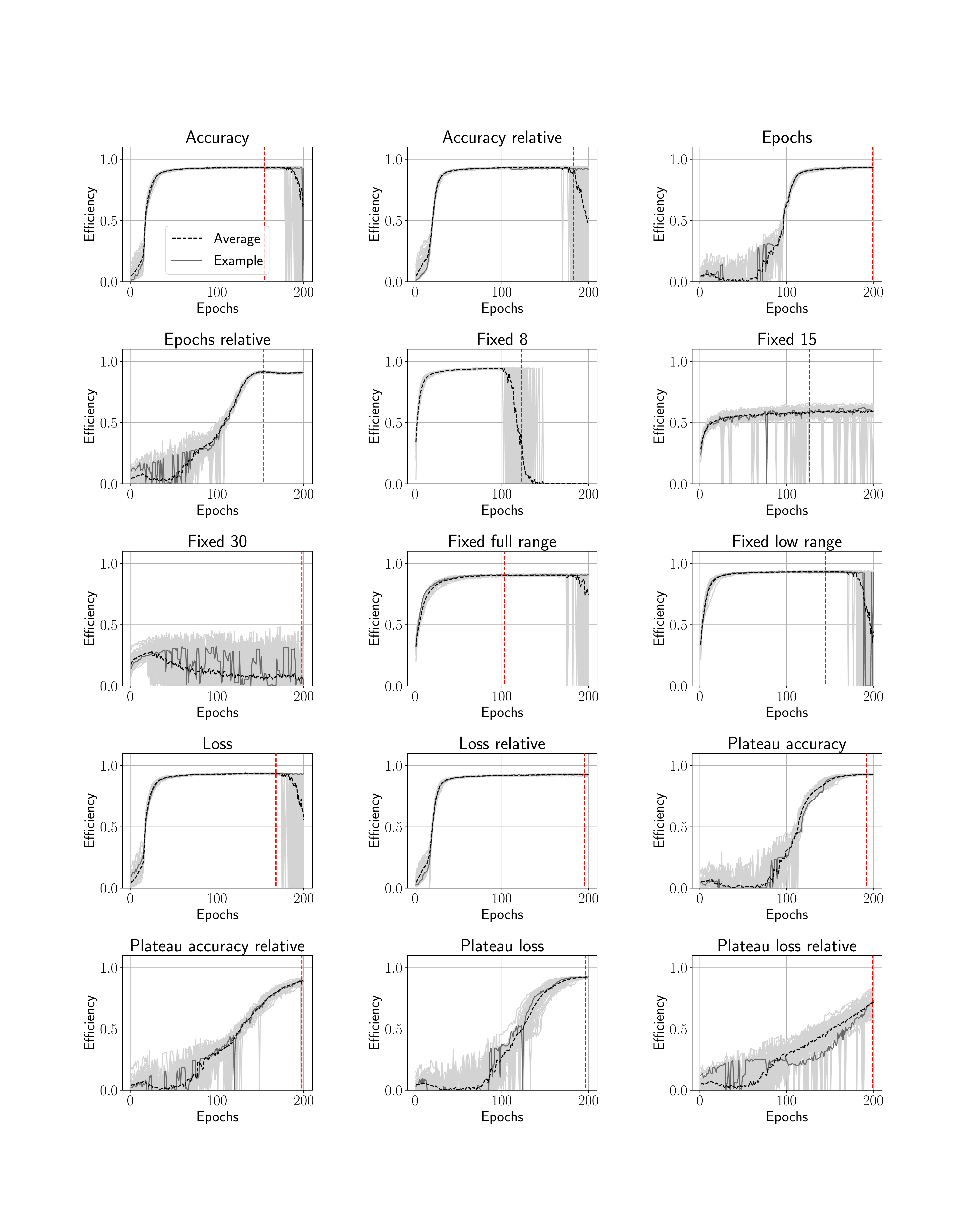}
    \caption{The efficiency for all $15$ tested training strategies as a function of the training epochs at \gls{snr} $9$ and a \gls{fap} of $10^{-4}$. The light grey curves show the efficiency for the $50$ independent initialized training runs. The black dashed line shows the average over these individual runs. We highlight the evolution of a single run in dark grey. The vertical, red, dashed line signifies the epoch with the largest efficiency. We choose $3$ networks at this epoch for which to calculate the sensitive volume.}
    \label{fig:efficiency_evolution_all}
\end{figure*}

We trained $50$ networks for every training strategy discussed in \autoref{sec:methods:training_strategies}. \autoref{fig:efficiency_evolution_all} shows the evolution of the efficiency at \gls{snr} 9 for every training strategy. The networks use a Softmax activation on the final layer. While we also monitor different \gls{snr}s, it is this region we are most interested in for three reasons. The first is practical in nature. Above an \gls{snr} of $9$ networks trained with almost all training strategies recover close to $100\%$ of the signals. It is, therefore, impossible to separate them by efficiency. Secondly, most \gls{gw}s are expected to be detected at low \gls{snr}s \cite{Schutz:2011}. Hence, efficiency at low \gls{snr}s is most important. Lastly, \gls{snr} 8 is often used as a threshold above which matched-filter searches can comfortably detect most signals (compare \autoref{fig:sensitivity_fixed_low}). By probing the efficiency close to this threshold we can get a sense of how well the search is doing overall.

Most training strategies do not have a major impact on the maximum efficiency. With the exception of training with a fixed \gls{snr} $15$ and $30$ all converged networks reach efficiencies of $90\%$ (''Fixed full range'') to $94\%$ (''Fixed 8''). At \gls{snr} $6$ the efficiency consistently drops to $24\%$ (''Fixed full range'') to $33\%$ (''Loss relative'') for all converged networks other than the above mentioned exceptions. Above an \gls{snr} of $12$ the efficiencies reach $100\%$ for all networks except ''Fixed 15'' and ''Fixed 30''. Those only achieve $100\%$ efficiency at \gls{snr} $15$ and $21$ respectively.

The relative plateau strategies did not manage to converge within the first $200$ epochs. For these, we have extended the training length to $400$ epochs, which has allowed these runs to converge. They have reached comparable efficiencies to those mentioned above.

The main difference between all runs is the number of epochs required to reach a converged state. One can see that strategies which supply low \gls{snr} signals earlier reach their maximal efficiency earlier. This is especially emphasized with the runs ''Fixed 8'', ''Fixed full range'' and ''Fixed low range''. The curriculum strategies that use the accuracy or loss as their condition also converge quickly. They too supply low \gls{snr} samples very early on, as the respective condition is fulfilled at each of the first few epochs. Waiting for a set number of epochs to pass hinders the ability of the network to see low \gls{snr} signals early on and, therefore, takes more time to converge. The slowest converging strategies wait for the loss or accuracy to stop improving. They effectively have to wait at least $6$ epochs before lowering the training range. Using a relative approach to lowering the \gls{snr} range further decreases the speed at which low \gls{snr}s are explored. In the most extreme cases the networks do not converge within the given number of epochs.

Finally, some training strategies become unstable toward the end. All of these unstable strategies converge relatively fast. This suggests that the longer one trains a converged network the more likely the efficiency is to collapse. We, therefore, expect that strategies where the efficiency did not collapse during the first $200$ epochs would see a similar problem during later epochs.

\begin{figure*}
    \centering
    \includegraphics[width=\textwidth, trim=2.5cm 5cm 2.5cm 5cm]{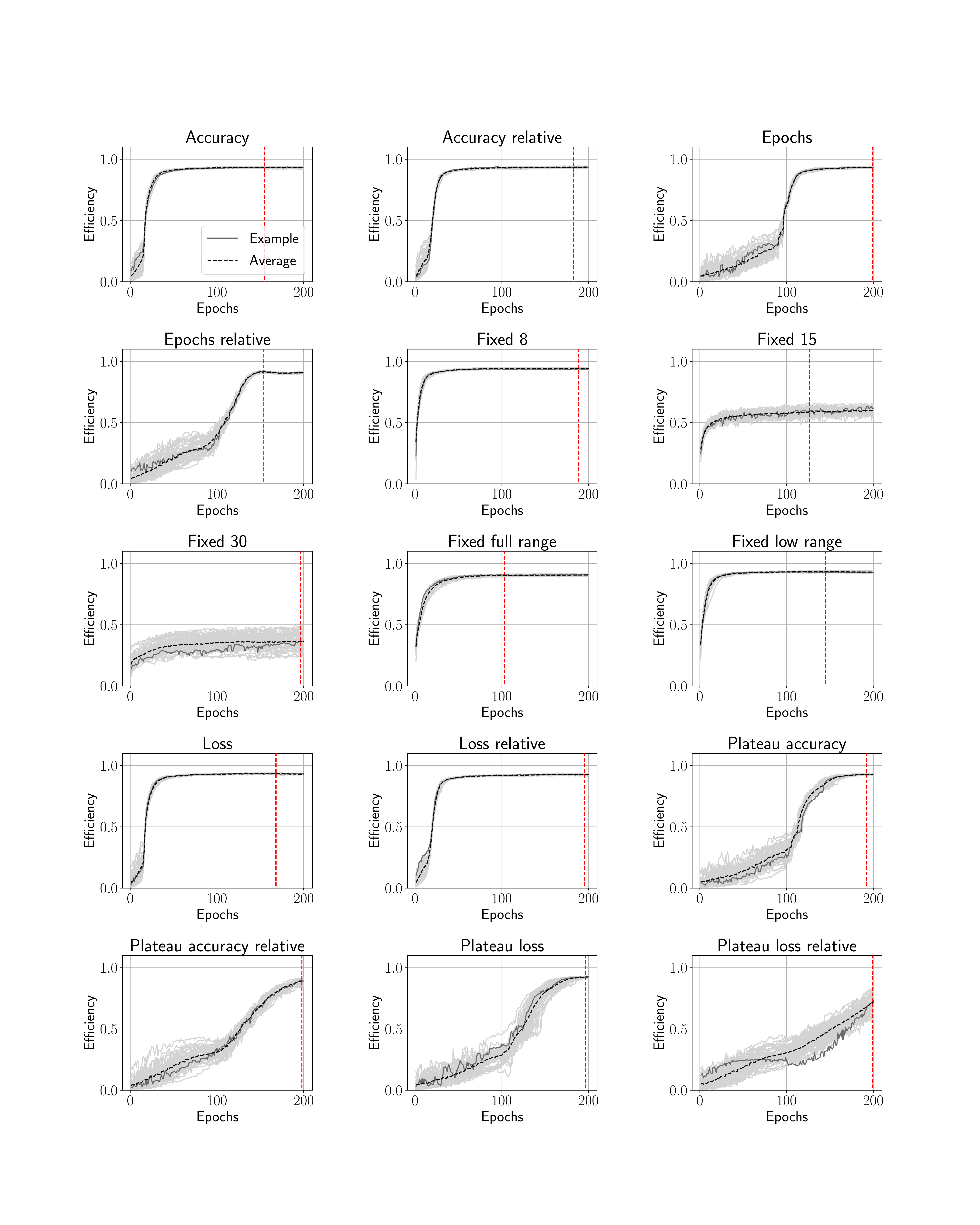}
    \caption{The efficiency for all $15$ tested training strategies as a function of the training epochs at \gls{snr} $9$ and a \gls{fap} of $10^{-4}$. The light grey curves show the efficiency for the $50$ independent initialized training runs. The black dashed line shows the average over these individual runs. We highlight the evolution of a single run in dark grey. The vertical, red, dashed line signifies the epoch with the largest efficiency. We choose $3$ networks at this epoch for which to calculate the sensitive volume. This figure shows the same networks as \autoref{fig:efficiency_evolution_all} with the \gls{usr} modification applied. With this modification the efficiency stays $>0$ at all times.}
    \label{fig:efficiency_evolution_all_lin}
\end{figure*}

The breakdown of the efficiency was resolved by the \gls{usr} modification in \autoref{sec:results:sensitivities}. \autoref{fig:efficiency_evolution_all_lin} shows the evolution of the efficiency at \gls{snr} $9$ when this fix is applied. We find that the drops to zero efficiency are removed but the qualitative features of the efficiency curves stay the same.

All efficiency plots were generated from the TensorFlow version of the networks. When training with PyTorch we found the results virtually indistinguishable from the TensorFlow version.

We repeated our tests on networks with different capacities, although this was not the main focus of the present work, to ensure that our findings are robust against a few specific architecture changes. We found no significant differences in the final efficiency, although the speed of training convergence varied. Such studies are left to future work.

\section{Conclusions}\label{sec:conclusion}
In this paper, we revisited the first deep learning \gls{gw} search algorithms and compared them directly to a matched-filter search. We showed that for the considered parameter space and for a single detector the networks retain performance closely following matched filtering even on long duration continuous data sets and when considering \gls{far} thresholds down to once per month. While there are now more sophisticated deep learning algorithms available that enhance the capabilities of the first proofs of concept, we think that there is still a lot to be learned from these first steps.

Our initial focus was the optimization of the data presentation to these networks. Two kinds of training strategies were previously explored; curriculum learning, where training samples become more difficult to classify as training continues, and fixed interval training, where the complexity of the training set stays constant.

We found that the particular strategy is of little importance to the eventual performance of the network. It depends a lot more on the presence of sufficiently complex samples in the training set. In particular, we found that the networks are able to generalize low \gls{snr} signals to high \gls{snr} ones but not vice versa.

On the other hand, the training strategy does have an impact on the time it takes the network to converge. Since high \gls{snr} examples are not as important to the performance of the network, strategies that provide low \gls{snr} samples earlier converge faster. In conclusion, we recommend training deep learning search algorithms on a fixed range of low \gls{snr} signals.

We use efficiency as our metric of performance during training. As this statistic has been used in previous publications, it allows us to verify that we have converged to the expected performance.

The efficiency at \gls{fap}s $\leq 10^{-4}$ dropped to zero when networks were trained for extended periods of time. This was unexpected and limited our ability to test the search.

We found the drops in efficiency to be caused by numerical instabilities in the final activation function of the networks. By removing the Softmax activation on the final layer and imposing thresholds directly on the linear output of the network, we were able to lift the limitations on the testable \gls{fap}s. This \gls{usr} modification has proven to be simple and effective, as no re-training of the networks is required and virtually unlimited low \gls{fap}s can be tested.

To compare the deep learning based searches to an equivalent matched-filter search we calculated the sensitive volumes as functions of the \gls{far}s on a month of simulated data. We found that the machine learning algorithm is able to closely follow the performance of the traditional algorithm even down to \gls{far}s of 1 per month, when \gls{usr} is used.

The results given here are limited to a single detector, Gaussian noise and signals from \gls{bbh}s, which are relatively short in duration and comparatively simple to detect with existing methods. Parts of the parameter space, like the inclusion of higher-order modes \cite{Harry:2017:highermodes}, eccentricity \cite{Nitz:2019:eccentric}, or precession \cite{Harry:2016:precessing}, where current searches are computationally limited, are not yet included. However, it is expected that neural networks may generalize efficiently to these more difficult signals. Deep learning detection algorithms for spinning black holes with precession were recently explored for the first time by \cite{Wei2021}. There is also ongoing work to construct neural network searches targeting long duration signals \cite{Dreissigacker:2020, Krastev:2020, Schaefer:2020, Wei:2020}. Considering real noise may enable deep learning algorithms to outperform matched filtering, which is only known to be optimal for stationary Gaussian noise. Multiple studies have shown that neural networks adapt well to non-stationary noise contaminated with glitches \cite{Huerta:2018b,Wei:2020,Wei2021,Yan2021}.

\section{Acknowledgments}
We acknowledge the Max Planck Gesellschaft and the Atlas cluster computing team at Albert-Einstein Institut (AEI) Hannover for support, as well as the ARA cluster team at the URZ Jena. F.O. was supported by the Max Planck Society's Independent Research Group Programme. O.Z. thanks the Carl Zeiss Foundation for the financial support within the scope of the program line "Breakthroughs".

\appendix
\section{Efficiency curve examples}
This appendix provides examples of the usefulness of \gls{usr} for various training strategies explored in this paper. The plots show the efficiency as a function of training epochs at 4 distinct \gls{snr}s with and without the application of the \gls{usr}. For both shown examples the \gls{usr} manages to remove the efficiency breakdown entirely.

\begin{figure*}
    \centering
    \includegraphics[width=\textwidth]{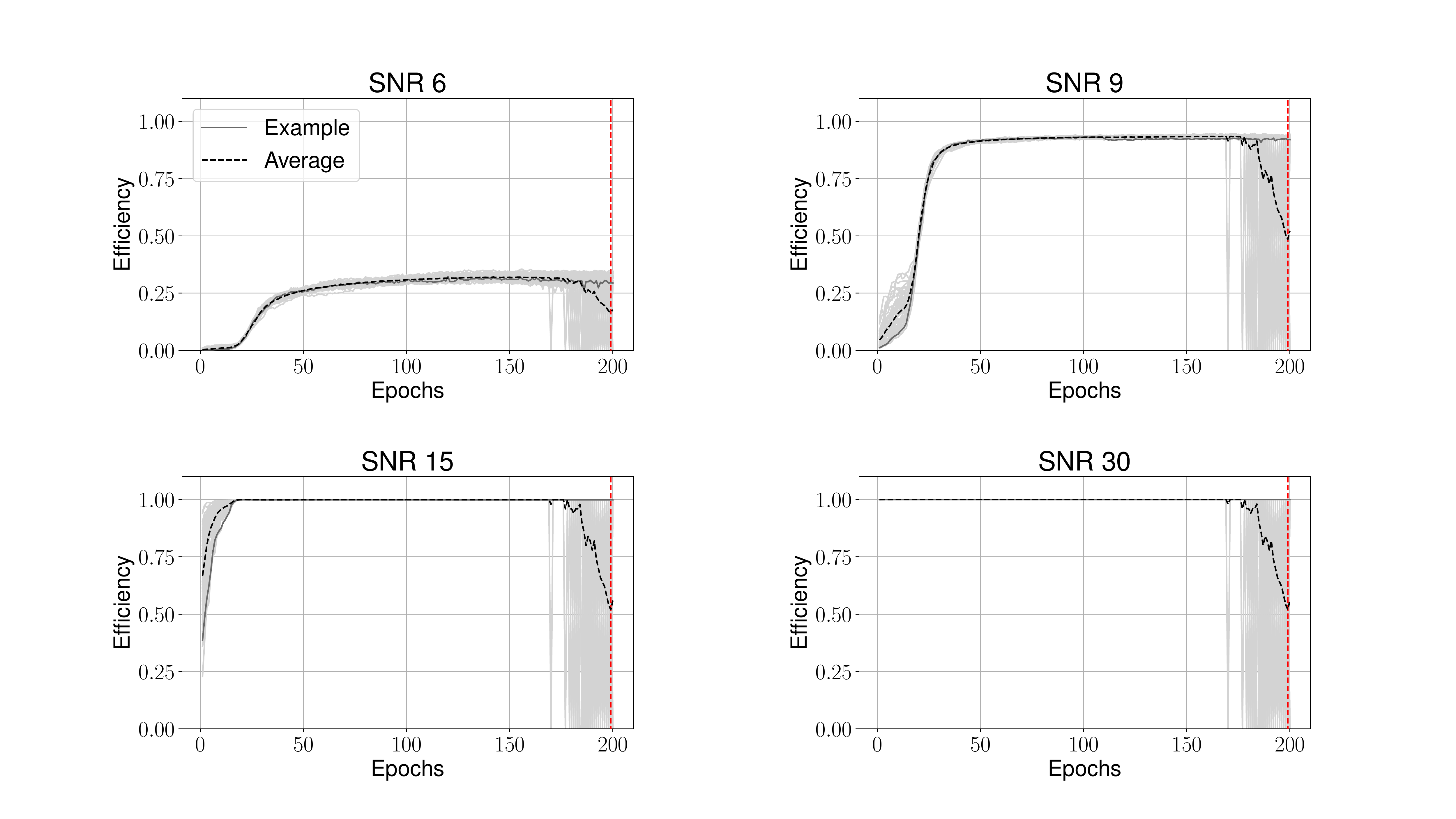}
    \caption{Efficiency evolution of the "Accuracy relative" strategy using the Softmax output as a ranking statistic.}
    \label{fig:efficiency_evolution_acc_rel_soft}
\end{figure*}

\begin{figure*}
    \centering
    \includegraphics[width=\textwidth]{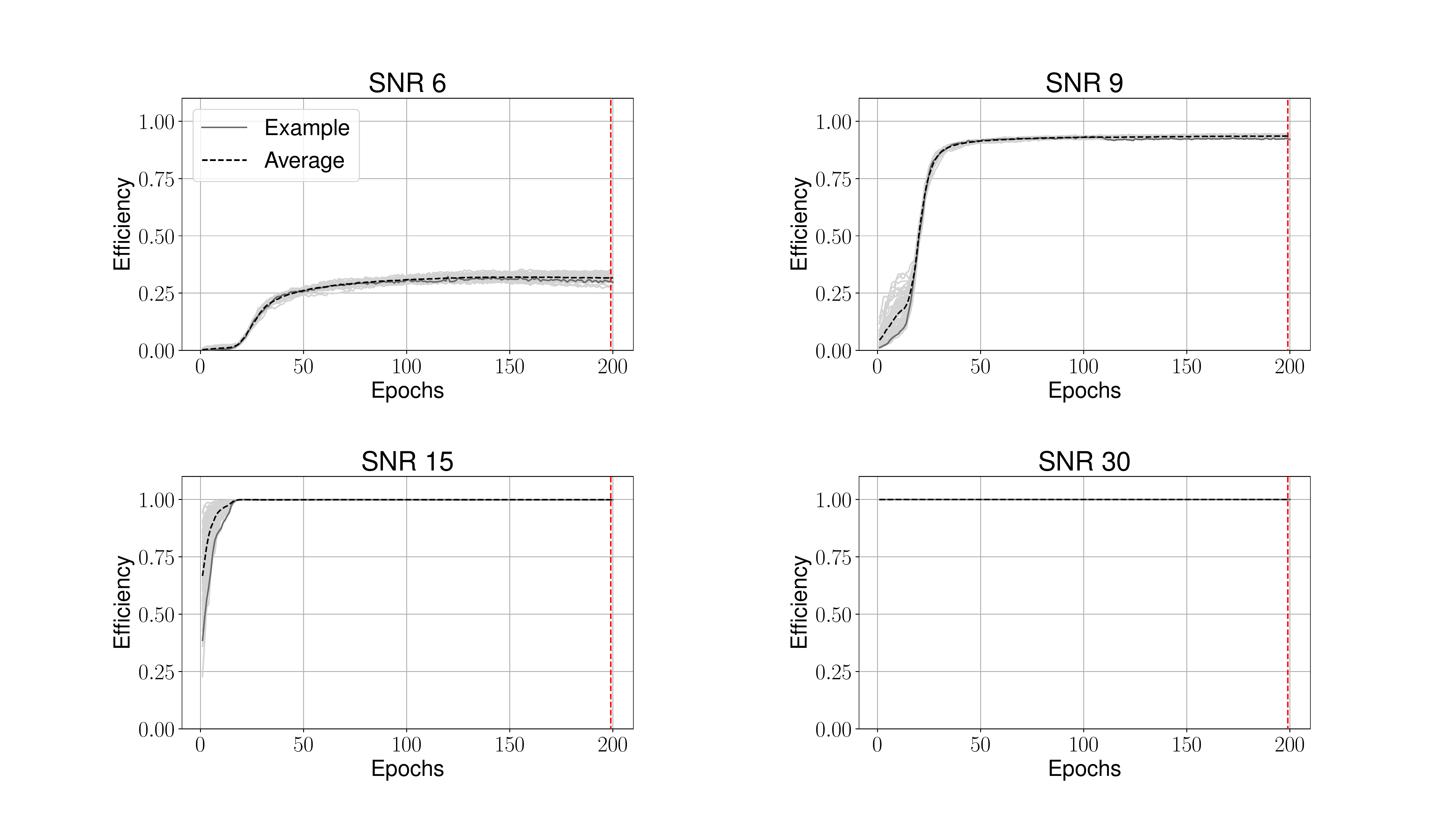}
    \caption{Efficiency evolution of the "Accuracy relative" strategy using the \gls{usr} modification.}
    \label{fig:efficiency_evolution_acc_rel_lin}
\end{figure*}

\begin{figure*}
    \centering
    \includegraphics[width=\textwidth]{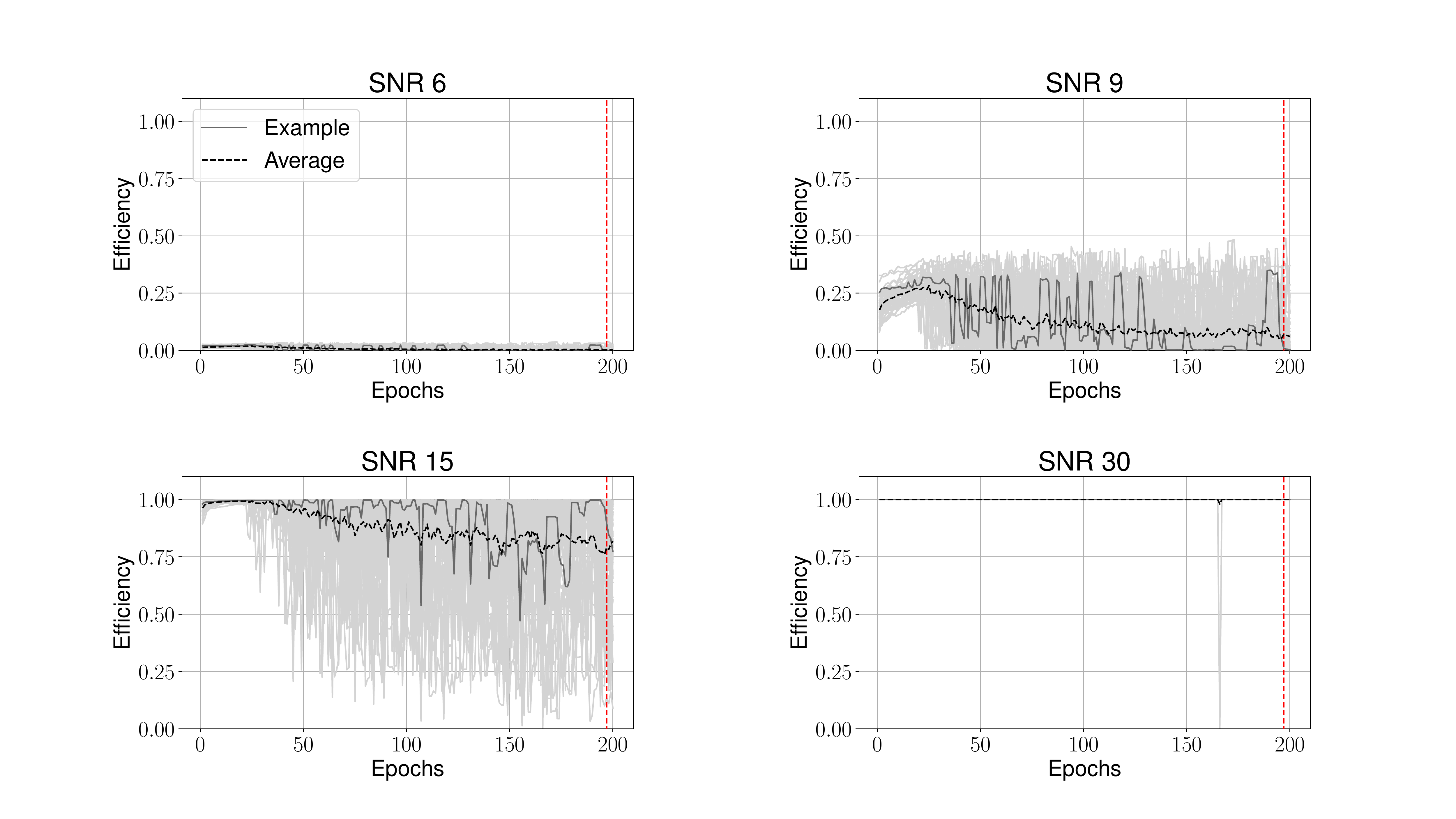}
    \caption{Efficiency evolution of the "Fixed 30" strategy using the Softmax output as a ranking statistic.}
    \label{fig:efficiency_evolution_fixed_30_soft}
\end{figure*}

\begin{figure*}
    \centering
    \includegraphics[width=\textwidth]{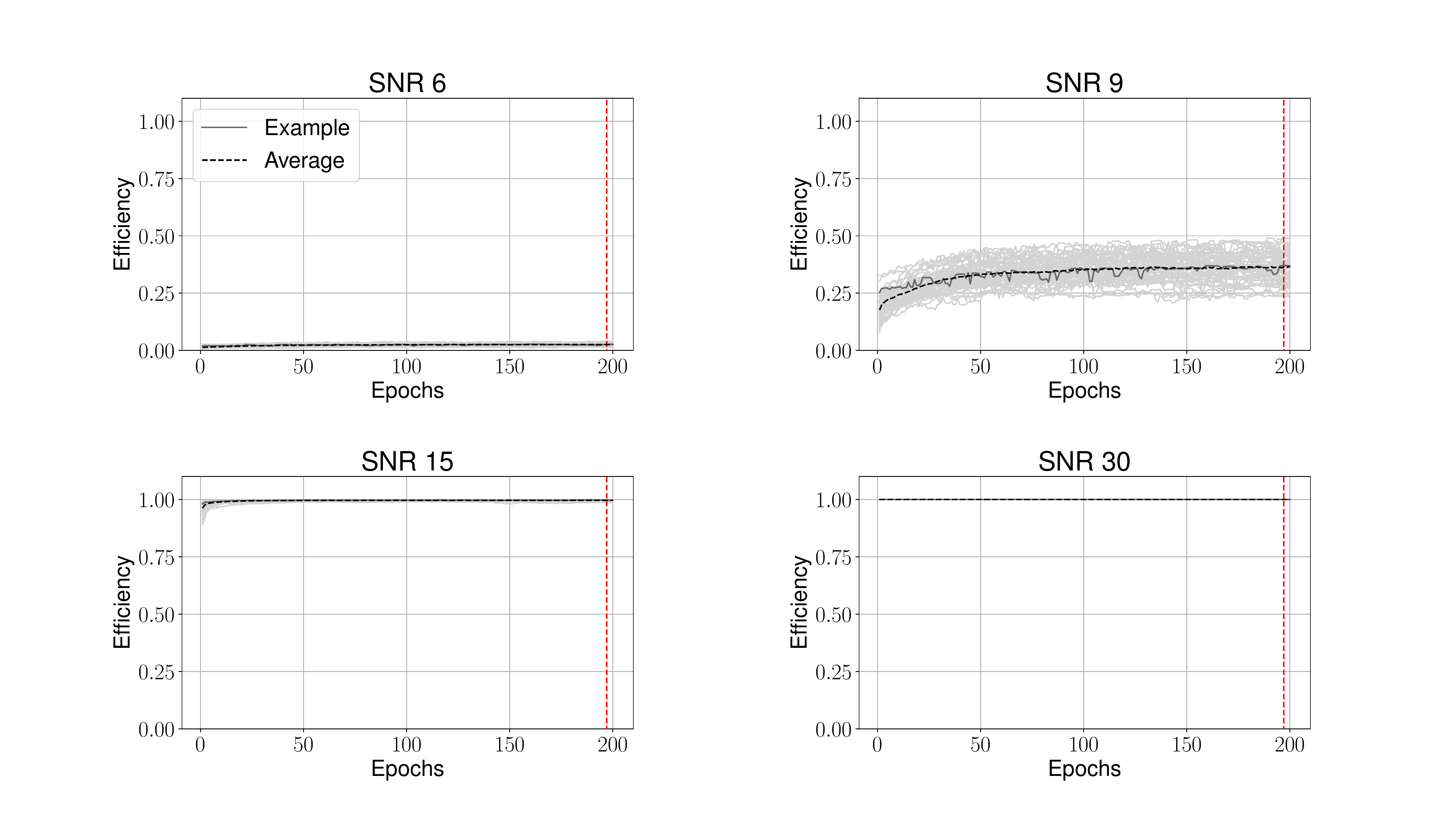}
    \caption{Efficiency evolution of the "Fixed 30" strategy using the \gls{usr} modification.}
    \label{fig:efficiency_evolution_fixed_30_lin}
\end{figure*}

\clearpage

\bibliography{bibliography}

\end{document}